\documentclass[aps,prl,twocolumn,superscriptaddress,nofootinbib]{revtex4-1}
\usepackage{amssymb,graphicx,color}
\usepackage[intlimits]{amsmath}
\usepackage[english]{babel}
\usepackage[colorlinks]{hyperref}

\hypersetup{
    colorlinks=red,
    linkcolor=red,
    filecolor=red,
    urlcolor=blue,
    citecolor=green
}

\usepackage[normalem]{ulem}
\usepackage{comment}
\usepackage{ragged2e}
\usepackage{enumerate}
\usepackage{subfigure}
\usepackage{float}

\begin{document}

\title{Phase transitions of the Dicke model: a unified perspective}

\author{Pragna Das}
\affiliation{Indian Institute of Science Education and Research Bhopal 462066 India}
\author{Auditya Sharma}
\affiliation{Indian Institute of Science Education and Research Bhopal 462066 India}

\begin{abstract}
  The Dicke model exhibits a variety of phase transitions.  The
  quantum phase transition from the normal phase to the super-radiant
  phase is marked by a dramatic change in the scaling of the
  participation ratio. We find that the ground state in the
  super-radiant phase exhibits multifractality manifest in the
  participation ratio scaling as the square root of the full Hilbert
  space dimension. The thermal phase transition temperature, for which
  we obtain an exact analytical expression, is strikingly captured by
  the mutual information between two spins. In the excited state
  quantum phase transition within the super-radiant phase, we discover
  a new upper cut-off energy; the central energy band between the
  lower and upper cut-off energies shows distinctly different
  behaviour. This finding is corroborated with the aid of several
  \emph{eigenvector} properties: von Neumann entanglement entropy
  between spins and bosons, the mean photon number, concurrence
  between two spins, and participation ratio. Thus we obtain a unified
  picture for the three different kinds of phase transitions. 
\end{abstract}

\maketitle 

The Dicke model, which incorporates the interactions of an ensemble of
$N$ two-level atoms via dipole coupling with a single bosonic
mode~\cite{ haroche2007short, kimble1996cavity, mirhosseini2019cavity}
has its origin in quantum optics, but has found application in a wide
range of fields from quantum chaos to quantum entanglement~\cite{
  furuya1998quantum, lakshminarayan2001entangling,
  bandyopadhyay2002testing, bandyopadhyay2004entanglement,
  tanaka2002saturation, vznidarivc2005generation,
  jacquod2004semiclassical, ghose2004entanglement,
  demkowicz2004global, weinstein2005entanglement,
  lakshminarayan2003entanglement,lambert2004entanglement,
  emary2003chaos, emary2003quantum} to scrambling and
thermalization~\cite{lewis2019unifying}. Besides possessing an
intimate connection to experiments~\cite{baumann2011exploring, 
klinder2015dynamical, baden2014realization}, the Dicke model is a
testbed for a variety of phase
transitions~\cite{kirton2019introduction}.  Although a lot is known
about these different transitions, the literature presents a rather
scattered treatment of them~\cite{dicke1954coherence,
  kadantseva1990superradiance, kirton2018superradiant,
  kirton2019introduction,hepp1973superradiant,wang1973phase,perez2017thermal,
  zhu2019entanglement}. In this Letter, we provide a transparent
unified picture of three different kinds of phase transitions in the
Dicke model.

The nature of the ground state of the Dicke model is dramatically
different depending on the magnitude of the coupling between the atoms
and the field. While for small coupling, in the normal phase, the
average photon number in the ground state is close to zero, when the
coupling is greater than a critical value, in the super-radiant phase,
the ground state mean photon number scales linearly with the number of
atoms~\cite{kirton2019introduction,dicke1954coherence,
  kadantseva1990superradiance, kirton2018superradiant}. Entanglement
properties~\cite{lambert2004entanglement, emary2003chaos,
  emary2003quantum} offer clear signatures of this quantum phase
transition~\cite{Sachdev_1998, sondhi1997continuous,
  cejnar2010quantum, casten2007quantum, ma2009fisher,
  osterloh2002scaling} (QPT). Furthermore, a study of level statistics
shows that the system in fact also undergoes a transition from
quasi-integrable to quantum chaotic~\cite{emary2003chaos} at the
QPT. In this Letter, with the aid of a careful study of the
participation ratio~\cite{tsukerman2017inverse} of the ground state,
we show how the normal to super-radiant phase transition is really a
localization-to-multifractal transition. We find that in the
super-radiant phase, the ground state participation ratio scales as
the square root of the full Hilbert space dimension.

The Dicke model also exhibits a thermal phase transition (TPT) which
was realized many decades
ago~\cite{hepp1973superradiant,wang1973phase}. When the coupling is
greater than the critical coupling, as the temperature is increased,
we see a transition back from the super-radiant to the normal
phase~\cite{perez2017thermal}. In this Letter, we obtain an exact
analytical expression for the transition temperature. Generalizing the
approach of Wang and Hieo~\cite{wang1973phase}, we write down the
partition function as a double integral.  The transition temperature
is identified to be the point at which the method of steepest descent
used to evaluate the integral in the thermodynamic limit breaks
down. Furthermore, just like entanglement in the ground state marks
the quantum phase transition, we show how the mutual information (MI)
between atoms offers a striking signature at the thermal phase
transition.

The Dicke model also exhibits an excited state quantum phase
transition (ESQPT), a term that is used to denote criticality in the
excited states of a quantum system~\cite{perez2017thermal,
  caprio2008excited, stransky2014excited, zhu2019entanglement,
  cejnar2008impact, garcia2017excited}. The ESQPT, which is a
generalization of the QPT, and is characterized by abrupt variations
of the energy and other excited state properties at a sharp critical
value of the energy~\cite{ cejnar2021excited} must be viewed in the
backdrop of the tremendous recent interest in the properties of
excited states~\cite{pal2010many, nandkishore2015many,
  alet2018many,karthik2007entanglement, caprio2008excited,
  perez2011excited, perez2011quantum, beugeling2015global} of quantum
systems. In the present Letter, we uncover how the ESQPT of the Dicke
model affects not only energy levels below a certain lower cut-off,
but also the top-lying energy levels above a second~\emph{upper}
cut-off, a feature that has apparently been hitherto unnoticed in the
literature~\cite{perez2011excited, lewis2019unifying}. Strikingly, in
contrast to prior studies, we are able to identify these features with
the aid of several \emph{eigenstate} properties: von Neumann
entanglement entropy (VNEE), the mean photon number, concurrence and
$PR$. Supporting evidence comes from eigenvalue properties like level
statistics~\cite{poilblanc1993poisson} and the consecutive level
spacing ratio~\cite{atas2013distribution} considering both the whole
and different parts of the spectrum.

\begin{figure*}[htbp]
  \includegraphics[width=0.43\textwidth]{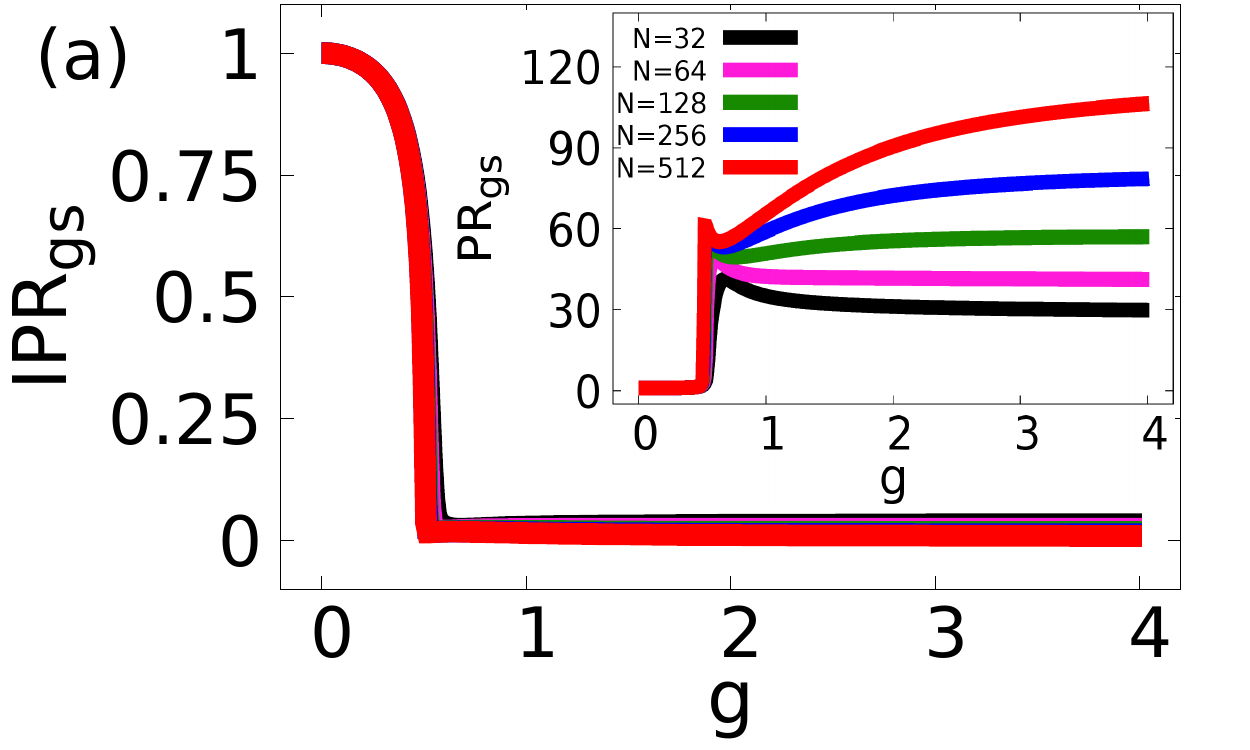}\
  \includegraphics[width=0.42\textwidth]{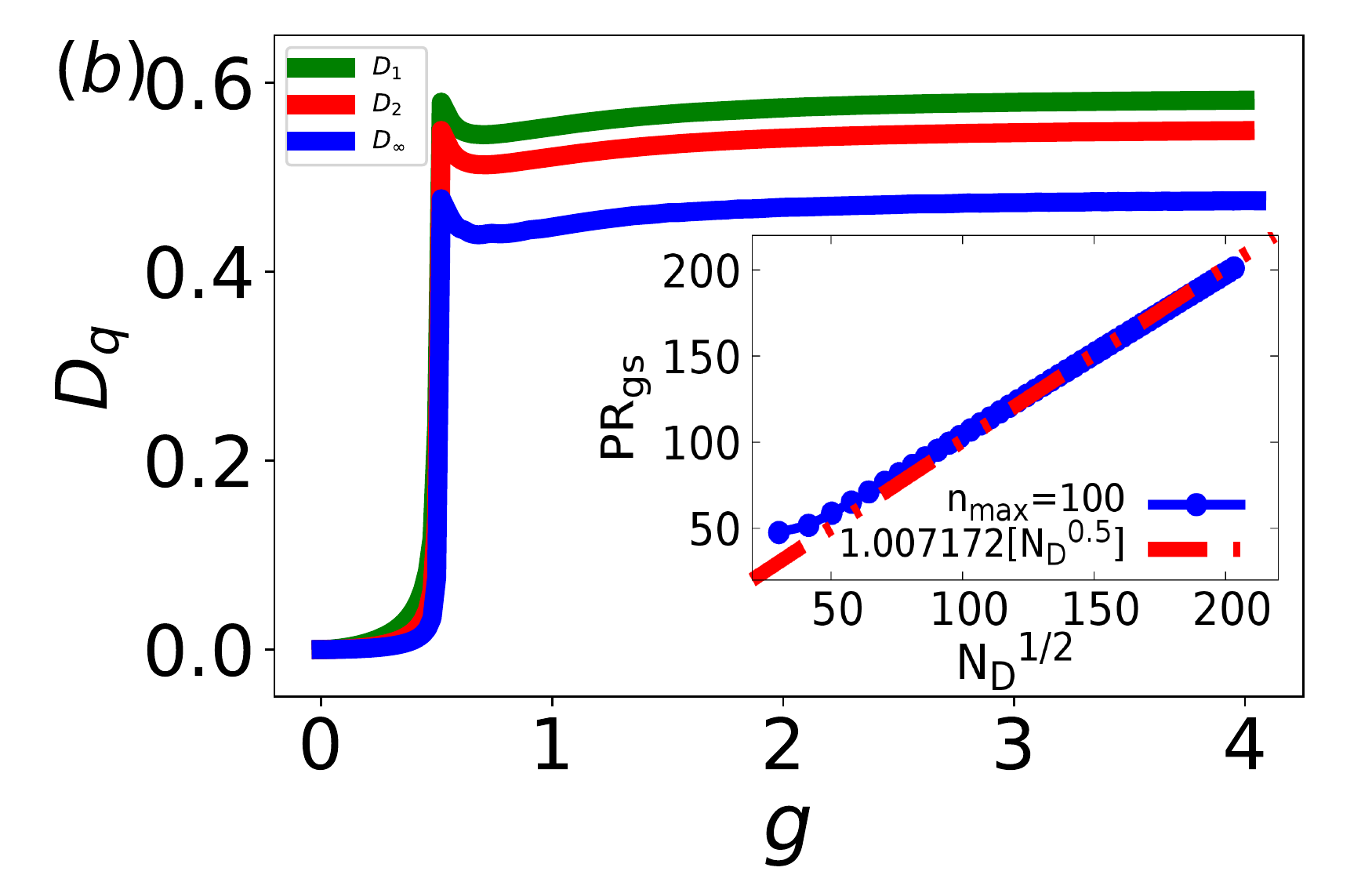}
  \caption{(a) The inverse participation ratio ($IPR$) of the ground
    state as a function of coupling $g$. The inset shows a similar
    plot for participation ratio ($PR$).  (b) Multifractal dimension
    $D_q$ ($q=1, 2, \infty$) of the ground state as a function of
    $g$. The inset shows the scaling of $PR$ with atom number $N$ for
    the ground state at $g=4.0$. In all the figures, we set $N=512$,
    $n_{\text{max}}=32$. }
  \label{fig:qpt2}
\end{figure*}

The Hamiltonian of the Dicke model is
\begin{eqnarray}
  \mathcal{H} &=& \omega a^{\dagger}a + \omega_0 J_z + \frac{g}{\sqrt{2j}}(a + a^{\dagger})(J_+ + J_-)
\end{eqnarray} 
where $a$ and $a^{\dagger}$ are bosonic operators satisfying the
commutation relation: $[a,a^{\dagger}]=1$ in units where $\hbar=1$.
$\omega$ is the single-mode frequency of the bosonic field while
$\omega_0$ is the level splitting of the atoms, and $g$ is the
coupling strength of the light-matter interaction. The angular
momentum operators
$J_{\pm,z}=\sum_{i=1}^{2j}\frac{1}{2}\sigma_{\pm,z}^{(i)}$ correspond
to a pseudospin with length $j$, composed of $N=2j$ spin-$\frac{1}{2}$
atoms described by Pauli matrices $\sigma_{\pm,z}^{(i)}$ acting on
site $i$ and satisfy the commutation relations: $[J_z,J_{\pm}]=\pm
J_{\pm}$, $[J_+,J_-]=2J_z$. The basis of the full Hilbert space of the
system is $\{\vert n\rangle\otimes\vert j,m\rangle\}$ where $\vert
n\rangle$ are the bosonic basis states satisfying $a^{\dagger}a\vert
n\rangle=n\vert n\rangle$ and $\vert j,m\rangle$ are the Dicke states
satisfying $J_{\pm}\vert j,m\rangle=\sqrt{j(j+1)-m(m\pm 1)}\vert
j,m\pm 1\rangle$, $J_z\vert j,m \rangle = m\vert j,m\rangle$. In our
work, we take $N$ to be even, and consider the symmetric subspace
which fixes $j = \frac{N}{2}$, and thus $m$ takes the $(N+1)$ values
$(-\frac{N}{2},...,0,...,\frac{N}{2})$. We also truncate the bosonic
mode to take the values $n = 0,1,...,n_{\text{max}}$. Thus the
dimension of the Hilbert space is given by $N_D =
(N+1)(n_{\text{max}}+1)$. In the thermodynamic limit the system shows
a second-order quantum phase transition from the normal phase (NP) to
the super-radiant phase (SP) at $g = \frac{\sqrt{\omega \omega_0}}{2}$
($=g_c$)~\cite{emary2003chaos}.  In all our numerical calculations we
have set $\omega=\omega_0=1$ and hence $g_c=0.5$.

{\it Quantum phase transition:} The inverse participation ratio ($IPR$) of an eigenstate
$\vert\psi\rangle=\sum_j^{N_D}\psi_j\vert j\rangle$ (where $N_D$ is
the Hilbert space dimension) defined as:
\begin{equation}
  IPR = \sum_{j=1}^{N_D}\vert\psi_j\vert^{4}.
\end{equation}
It is useful to quantify the degree of delocalization of the
eigenstate. Fig.~\ref{fig:qpt2}(a) shows an exact diagonalization
study of the $IPR$ of the ground state as $g$ is varied; we observe
that it is close to one in the NP and close to zero in the SP. Thus we
see that the ground state is localized in the NP whereas it is
extended in nature in the SP. Echoes of these features are also found
in both static and dynamical studies of a variety of other measures of
quantum correlations (see supplementary section). The inset of
Fig.~\ref{fig:qpt2}(a) studies the participation ratio $PR$ (which is
the inverse of $IPR$), as a function of the coupling $g$ and it shows
a phase transition from NP where it takes values close to zero, to SP
with a sharp transition to a non-zero value at the critical coupling.
We plot the $PR$ for different atom number $N$ and notice that in the
SP the value of $PR$ increases with $N$.

A finer understanding of the localization properties may be obtained
by studying the multifractal dimension~\cite{mace2019multifractal,
  lindinger2019many}:
\begin{eqnarray}
  D_q &=& \frac{S_q}{\ln(N_D)}
\end{eqnarray}
where $S_q = \frac{1}{1-q}\ln\Big
(\sum_{j=1}^{N_D}\vert\psi_j\vert^{2q}\Big)$ is known as the
$q$-dependent participation entropy. In the Shannon limit ($q=1$),
$S_1 = \sum_j\vert\psi_j\vert^{2}\ln\Big(\vert\psi_j\vert^{2}\Big)$,
while $q=2$ yields the usual $IPR$ with $S_2=-\ln(IPR)$.  $S_{\infty}$
is determined by the maximum value of the densities
$p_{\text{max}}=\text{max}_j\vert\psi_j\vert^{2}$ and $D_{\infty}=
-\frac{\ln(p_{\text{max}})}{\ln(N_D)}$. For a perfectly delocalized
state $S_q=\ln(N_D)$ (when $N_D$ is large) and hence $D_q=1$ for all
$q$. On the other hand for a localized state $S_q=constant$ and
$D_q=0$. In an intermediate situation, wave functions are extended but
non-ergodic with $S_q=D_q\ln(N_D)$ where $0<D_q<1$ and the state is
multifractal. In Fig.~\ref{fig:qpt2}(b) we show $D_1$, $D_2$ and
$D_{\infty}$ for the ground state as a function of the coupling
parameter $g$. In the NP $D_q \approx 0$ hence we can say that the
ground state is localized in the NP.  Contrastingly in the SP,
$0<D_q<1$ with $D_1>D_2>D_{\infty}$ ($D_1\approx 0.58$, $D_2\approx
0.55$, $D_{\infty}\approx 0.47$), with a sharp transition at the
critical point. In the inset of Fig.~\ref{fig:qpt2}(b) we show that at
$g=4.0$, $PR$ goes as the square root of the Hilbert space
dimension. Hence we find that the SP is neither perfectly delocalized
nor localized, and in fact displays multifractal
character~\cite{mace2019multifractal}. Given the intense current
interest in multifractal states~\cite{lindinger2017multifractal, 
pino2017multifractal, serbyn2017thouless}, this discovery in a
familiar model is an exciting finding.

\begin{figure*}[htbp]
  \includegraphics[width=0.345\textwidth]{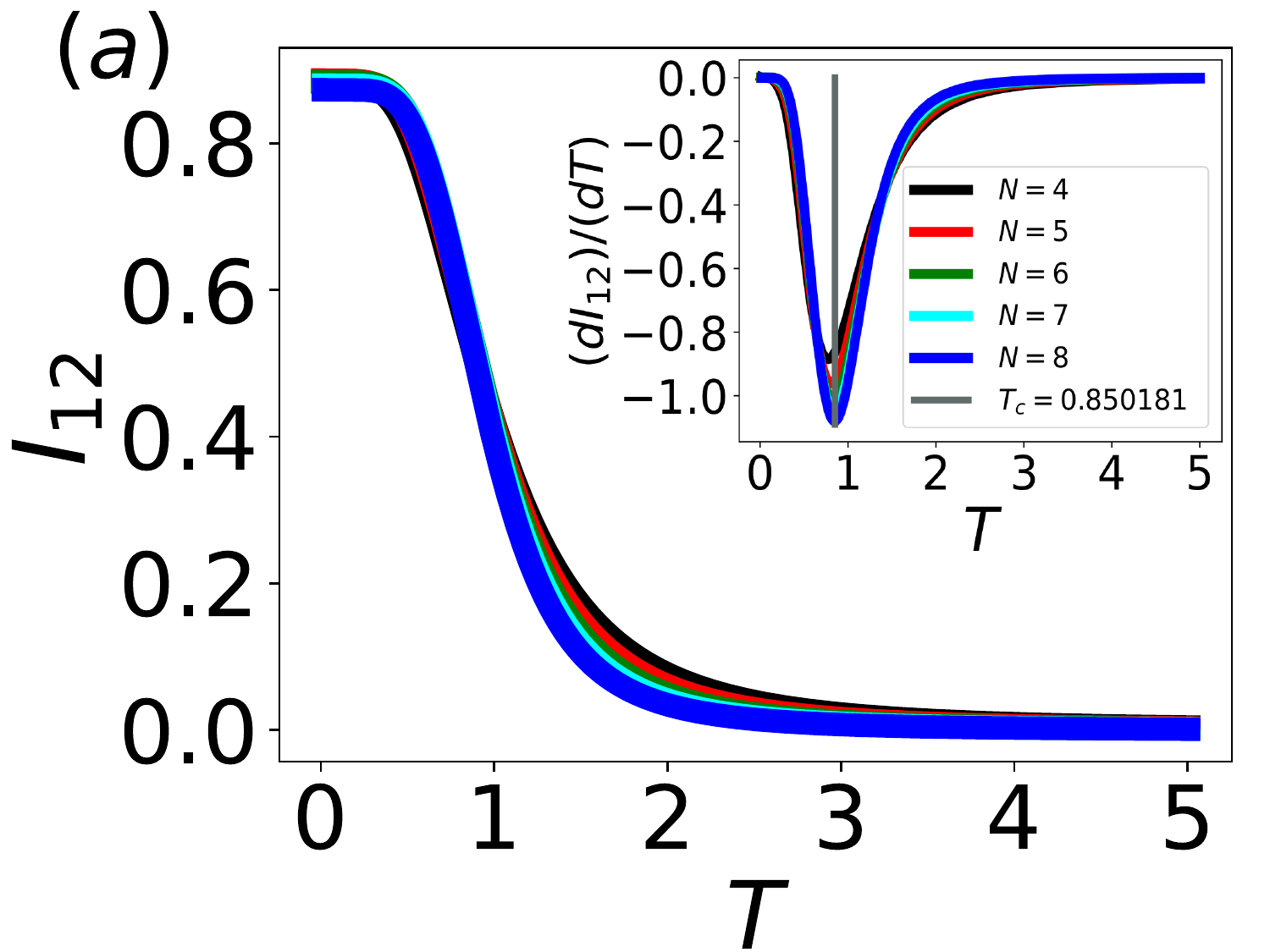}
  \includegraphics[width=0.325\textwidth]{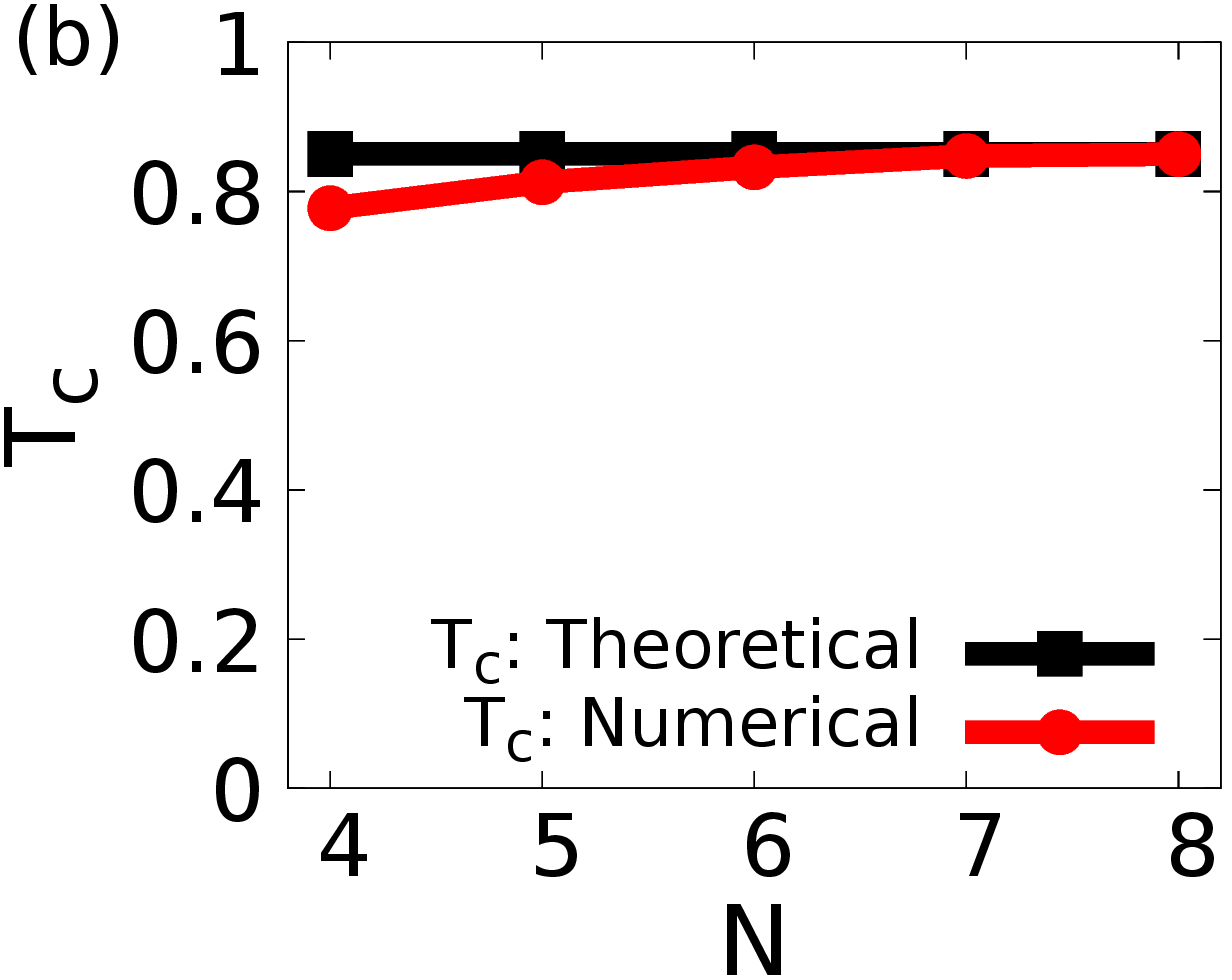}
  \includegraphics[width=0.31\textwidth]{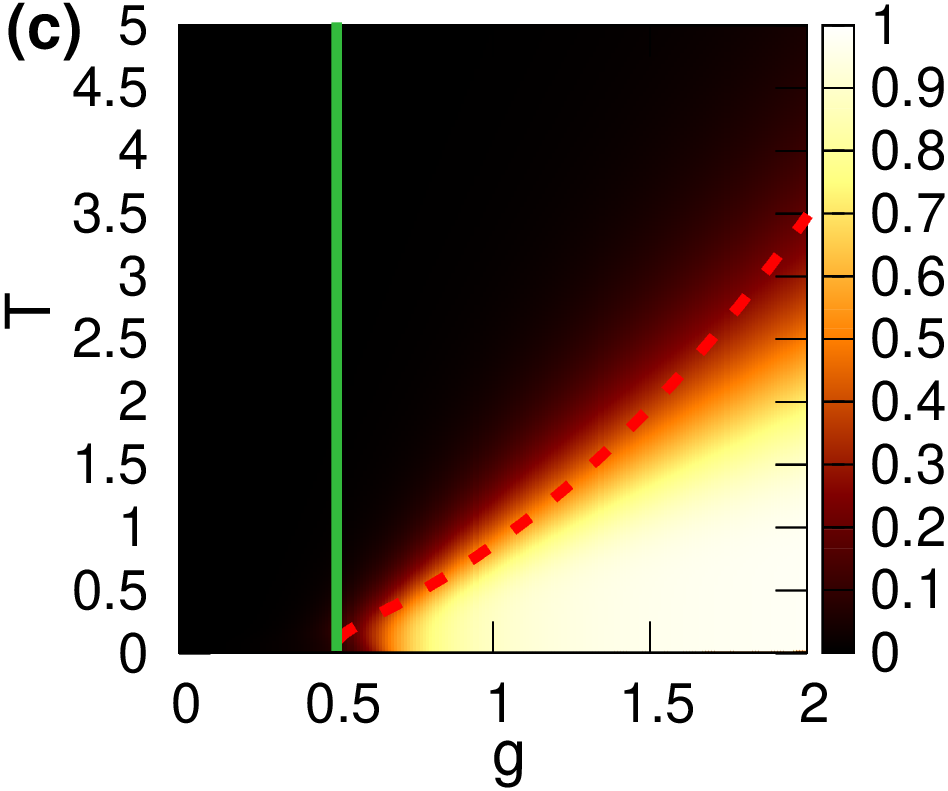}
  \caption{(a) Mutual information (MI) of two spins as a function of
    temperature at $g=1.0$, with the inset showing the numerical
    derivative of MI wrt temperature $\frac{dI_{12}}{dT}$. (b) The red
    line denotes the critical temperature as a function of atom number
    $N$, while the black line is the theoretical value
    $(T_c=0.850181)$ for $g=1.0$. (c) Mutual information of two spins
    as a function of coupling $g$ and temperature $T$. The parameters
    are $N=6$, $n_{max}=10$. The black region corresponds to the NP,
    and the white region to the SP. The solid line corresponds to
    $g_c$ and the dashed line denotes the critical temperature
    theoretically calculated in Eqn.~\ref{eqn:Tc}. In all the plots
    $\omega=\omega_0=1$.}
  \label{fig:tpt1}
\end{figure*}

{\it Thermal phase transition:} To compute the partition function
($Z = Tr(e^{-\frac{\mathcal{H}}{k_B T}})$) of the Dicke Hamiltonian it
is useful to first write it in units of $\omega$ as:
\begin{align}
  \tilde{\mathcal{H}} = \frac{\mathcal{H}}{\omega} = a^{\dagger}a + \sum_{j=1}^{N}\frac{\epsilon}{2}\sigma_j^z + \frac{\lambda}{\sqrt{N}}\sum_{j=1}^{N}( a + a^{\dagger} )\sigma^x
\end{align}
where $\epsilon = \frac{\omega_0}{\omega}$ and
$\lambda=\frac{g}{\omega}$. Following the method of Wang and
Hieo~\cite{wang1973phase} (who studied the Dicke model within the
rotating wave approximation), the computation of the partition function reduces to the evaluation of a double integral (see supplementary section): 
\begin{align}
  Z(N,T) = \int\frac{d^2\alpha}{\pi}e^{-\beta\vert\alpha\vert^2}\Big(2\cosh\Big[ \frac{\beta\epsilon}{2}\Big[ 1 + \frac{16{\lambda}^2\alpha^2}{\epsilon^2 N} \Big]^{1/2} \Big]\Big)^N
\end{align}
which in the thermodynamic limit ($N\to\infty$), may be carried out
with the aid of the method of steepest descent, within the
super-radiant phase. Tracking the point at which the method breaks
down (see supplementary section), we obtain an exact expression for
the transition temperature:
\begin{equation}
  T_c = \frac{1}{\beta_c} = \Big(\frac{\omega_0}{2\omega}\Big)\frac{1}{\tanh^{-1}\Big( \frac{\omega\omega_0}{4 g^2} \Big)}.
  \label{eqn:Tc}
\end{equation}
The critical temperature expression is meaningful only when
$g>g_c$. When $g < g_c$, the system is in the normal phase at all
temperatures.  When $g>g_c$, it is only above the critical temperature
that the system is in the normal phase, while for $T<T_c$ the system
is in the super-radiant phase.

\begin{figure*}[htbp]
  \includegraphics[width=0.24\textwidth]{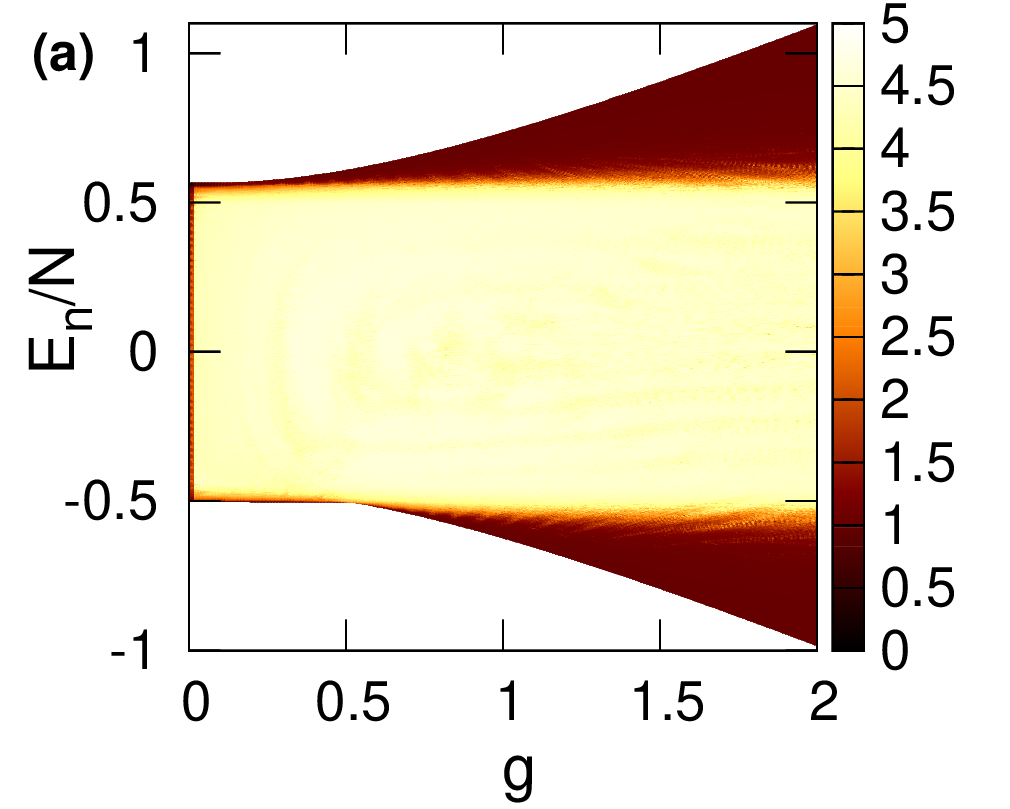}
  \includegraphics[width=0.24\textwidth]{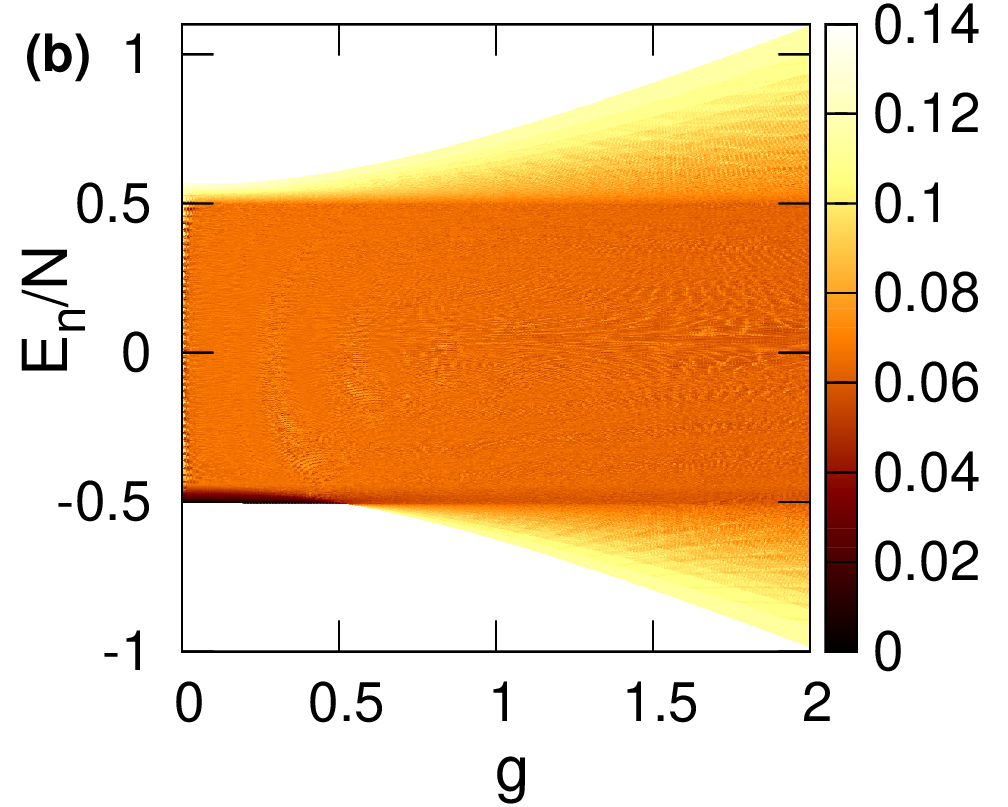}
  \includegraphics[width=0.24\textwidth]{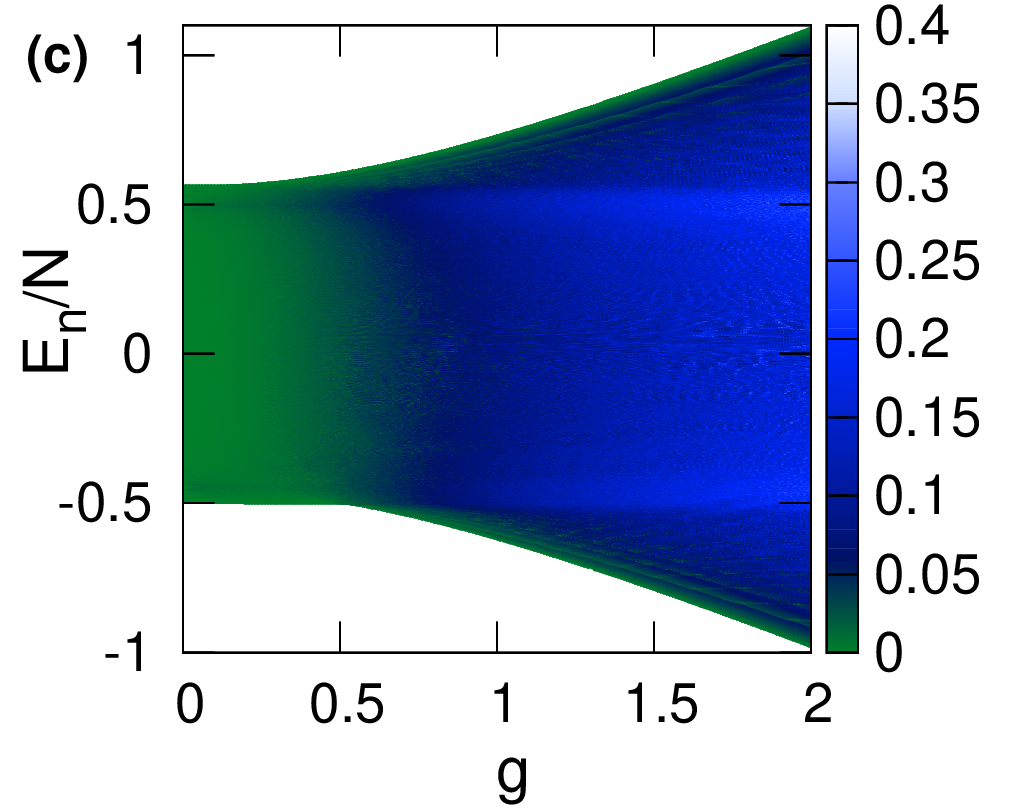}
  \includegraphics[width=0.24\textwidth]{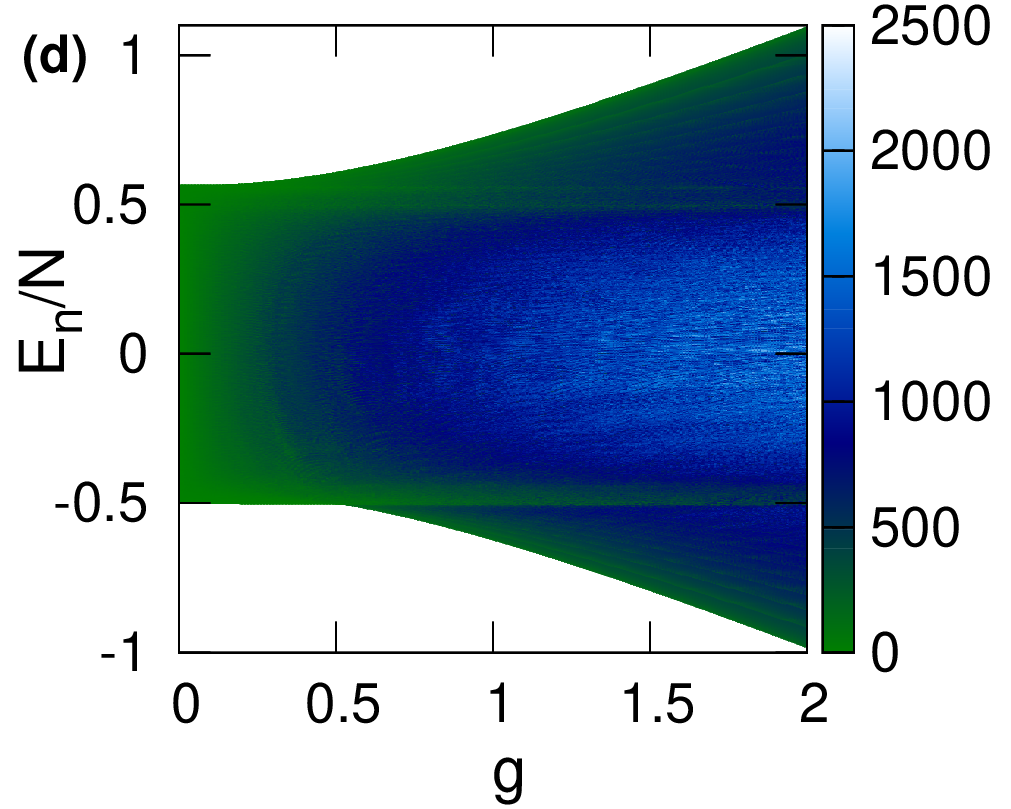}
  \caption{(a) Von Neumann entanglement entropy between the atoms and
    the bosons as a function of coupling strength $g$ and the energy
    density $E_n/N$ (eigenenergy divided by the atom number $N$) of
    the DM. (b) Similar plot for mean photon number, (c) for
    concurrence between any two atoms and (d) participation ratio of
    the eigenstates. The parameters for all
    the four plots are $N=512$, $n_{\text{max}}=32$.}
  \label{fig:esqpt1}
\end{figure*}

The mutual information (MI) between two atoms $I_{12}$ proves
to be a very useful quantity to study. Using the single-spin reduced density 
matrices $\rho_{1}$, $\rho_{2}$ and the two-spin reduced density matrix $\rho_{12}$, we can work out
the von Neumann entropies $S_{1,2}= -Tr(\rho_{1,2} \ln(\rho_{1,2}))$, $S_{12} = -Tr(\rho_{12}
\ln(\rho_{12}))$ from which the mutual information~\cite{vedral2003classical, 
divincenzo2004locking, adesso2010quantum, maziero2010quantum} is 
immediately written down:
\begin{equation}
  I_{12} = S_1 + S_2 - S_{12}.
\end{equation}
In Fig.~\ref{fig:tpt1}(a) we show $I_{12}$ as a function of
temperature at $g=1$ ($g>g_c$).  At low temperatures in the SP,
$I_{12}$ takes a value close to unity while at high temperatures in
the NP, $I_{12}$ drops to a value close to zero, with a dramatic drop
happening at a temperature close to the transition temperature. For a
finer understanding of the variation of the mutual information across
the transition, we study in the inset of Fig.~\ref{fig:tpt1}(a) the
first-order temperature derivative $\frac{d{I_{12}}}{dT}$, for
different atom numbers. We observe that the temperature at which the
derivative takes the minimum value is consistent with the transition
temperature $T_c$, denoted by the vertical
line. Fig.~\ref{fig:tpt1}(b) confirms that as the number of atoms is
increased the temperature at the minimum does indeed approach the
theoretically obtained critical temperature. From the surface plot of
the MI as a funtion of $g$ and $T$ in Fig.~\ref{fig:tpt1}(c), it is
clear that for $g<g_c$ there is no phase pransition, but for $g>g_c$
there exists a critical temperature $T_c$ at which the system changes
from the super-radiant phase ($T<T_c$) to the normal phase ($T>T_c$).
While it is widely known that entanglement in the ground state signals
the QPT, our work shows that despite also including classical
correlations, the mutual information between atoms offers a striking
signature at the thermal phase transition.

{\it ESQPT:} The Dicke model exhibits an excited state quantum phase
transition in the super-radiant phase. When $g> g_c$, it has been
reported~\cite{ perez2011excited, lewis2019unifying} that the
eigenvalues above a cut-off energy $E_c$ behave in a distinctly
different manner in comparison with the eigenvalues below the
cut-off. We find that in fact there is not just a lower cut-off, but
also an upper cut-off.  Our data show that we must study separately
the eigenvalues drawn from a central band that comprises of energy
levels between a lower and upper cut-off. The lower and upper energy
bands show different behaviour. While eigenvalue properties like
level-statistics and gap ratio provide supporting evidence (see
supplementary section), we highlight how \emph{eigenstate properties}
offer a striking demonstration of this picture.

In Fig.~\ref{fig:esqpt1}(a) we show the VNEE~\cite{lambert2004entanglement}
between spins and bosons:
\begin{equation}
S = - Tr\Big( \rho_{\text{boson}}\log_2\rho_{\text{boson}} \Big),
\end{equation}
for all the eigenstates of the Dicke model. Here $\rho_{\text{boson}}
= Tr_{\text{atom}}\rho$ is the reduced density matrix of the bosonic
part. We observe two cut-off energies: $(i)$ lower cut-off energy
(corresponding to the ground state energy at $g=g_c$) and $(ii)$ upper
cut-off energy (corresponding to the maximum energy for $g=0$).  The
value of VNEE is larger in the eigenstates of the central band in
comparison with that of the top and bottom bands. Thus the eigenstates
carry a clear signature of the two excited state quantum phase
transitions when $g > g_c$. In Fig.~\ref{fig:esqpt1}(b) we show a
similar plot for the mean photon number~\cite{emary2003chaos},
$\langle a^{\dagger}a\rangle$ which is scaled by the pseudospin length
$j$ of the system. It carries information pertaining to the bosonic
part of the eigenstates. In the middle band the value of the mean
photon number is comparatively lower than that of the other two
bands. However, we observe that neither the VNEE between the atoms and
the bosons, nor the mean photon number is able to distinguish between
the $g<g_c$ and $g>g_c$ regions of the middle band.  A study of the
entanglement between atoms provides useful further perspective.

The concurrence~\cite{hill1997entanglement, 
wootters1998entanglement, wootters2001entanglement, dennison2001entanglement} 
between (any) two atoms is given by: 
\begin{equation}
C = max \{0,\lambda_1 - \lambda_2 - \lambda_3 - \lambda_4 \},
\label{eqn:concurrence}
\end{equation}
where $\lambda_i$ are the square roots of the eigenvalues of the
matrix product, $\tilde{{\rho}}_{12} =
\rho_{12}(\sigma_{1y}\otimes\sigma_{2y})
\rho^{\star}_{12}(\sigma_{1y}\otimes\sigma_{2y})$, in decending order
($\lambda_1>\lambda_2>\lambda_3>\lambda_4$).  Here $\rho^{\star}_{12}$
denotes complex conjugation of $\rho_{12}$, and $\sigma_{iy}$ are
Pauli matrices for two-level systems. In Fig.~\ref{fig:esqpt1}(c) we
plot concurrence between two atoms for the whole spectrum as a
function of $g$. We observe that in addition to showing a signature of
the ESQPT in the super-radiant phase, concurrence is also able to
distinguish the eigenstates of the middle band in the $g<g_c$ region
and the $g>g_c$ region. In the NP, the concurrence value of the
central states is comparatively smaller than that for the central
states of the SP. Again the value of concurrence in the bottom and top
bands of the super-radiant phase is a bit lower than that of the
central band. Fig.~\ref{fig:esqpt1}(d) shows the participation ratio
of all the eigenstates as a function of coupling parameter $g$. We are
able to identify the ESQPT, which divides the whole spectrum into
three bands: top, bottom, and central. In the NP ($g<g_c$), the whole
region shows a uniform comparatively small value of $PR$. On the other
hand in the SP ($g>g_c$) while the central band exhibits a larger
value of $PR$, the top and bottom bands show mixed behaviour, although
they resemble the NP more than the SP.

{\it Summary:}\label{sec_5} We study the phase transitions (QPT, TPT,
ESQPT) of the Dicke model, with the aid of a number of measures of
localization, entanglement and mutual information. Different
quantities are more suitable for the different kinds of phase
transitions involved, and a comprehensive look at all of them helps us
obtain a unique overall big-picture of the Dicke model. We are thus
able to provide a unified perspective of three different kinds of
phase transitions. The $IPR$ for the ground state shows a sharp phase
transition at $g_c$; while in the NP the ground state behaves like a
localized state, the ground state in the SP is not localized. A
careful study of the scaling of $PR$ (for the ground state in the SP)
with the dimension $N_D$ of the full Hilbert space reveals that the
$PR$ scales as $\sqrt{N_D}$ suggesting multifractral character. In the
$g>g_c$ region there exists some critical temperature $T_c$, above
which the SP disappears and the system goes into the NP whereas for
$g<g_c$ the system remains in the NP for all temperatures. We obtain a
closed-form expression for the transition temperature in the
super-radiant phase, and numerically verify that the mutual
information between two atoms provides a useful signature at the
transition. Thus at the temperature transition, the mutual information
proves to be a worthy generalization of entanglement, which marks the
ground state QPT. We find that in the super-radiant phase, the ESQPT
is signalled not just by a lower energy cut-off, but an upper energy
cut-off as well. The ESQPT is studied with the aid of VNEE, mean
photon number, concurrence and $PR$. For the VNEE and mean photon
number the whole central band is uniform, with no distinction between
$g<g_c$ and $g>g_c$ regions. We find that concurrence and $PR$ reveal
more structure. In addition to showing a signature of the ESQPT in the
SP, these quantities are also able to distinguish the eigenstates of
the central band between the $g<g_c$ region and $g>g_c$ region. Hence
we are able to present various phase transitions in the DM in terms of
several quantities that measure localization, multifractality, mutual
information and entanglement. It would be interesting to extend the
ideas in this study to other spin-boson models, to open quantum
systems that include a bosonic bath, and models with a periodic drive.

\section*{Acknowledgments}
We are thankful to Devendra Singh Bhakuni, Nilanjan Roy, Suhas
Gangadharaiah and Sebastian W\"{u}ster for fruitful comments and
discussions. P.D. is grateful to IISERB for the PhD fellowship. A.S
acknowledges financial support from SERB via the grant (File Number:
CRG/2019/003447), and from DST via the DST-INSPIRE Faculty Award
[DST/INSPIRE/04/2014/002461].


%
\onecolumngrid

\end{document}


\title{Supplementary material for ``Multifractality, mutual information and entanglement in the Dicke model''}
	
\author{Pragna Das}
\affiliation{Indian Institute of Science Education and Research Bhopal 462066 India}
\author{Auditya Sharma}
\affiliation{Indian Institute of Science Education and Research Bhopal 462066 India}
	
\maketitle 

\section{Quantum phase transition}
The quantum phase transition in the Dicke model is profitably studied
with the aid of a number of quantities.  While the most characteristic
ones are presented in the main Letter, we show some useful ancillary
quantities here in the supplementary section.

\subsection{Statics}
    \begin{figure*}[htbp]
      \subfigure{\includegraphics[width=0.32\textwidth]{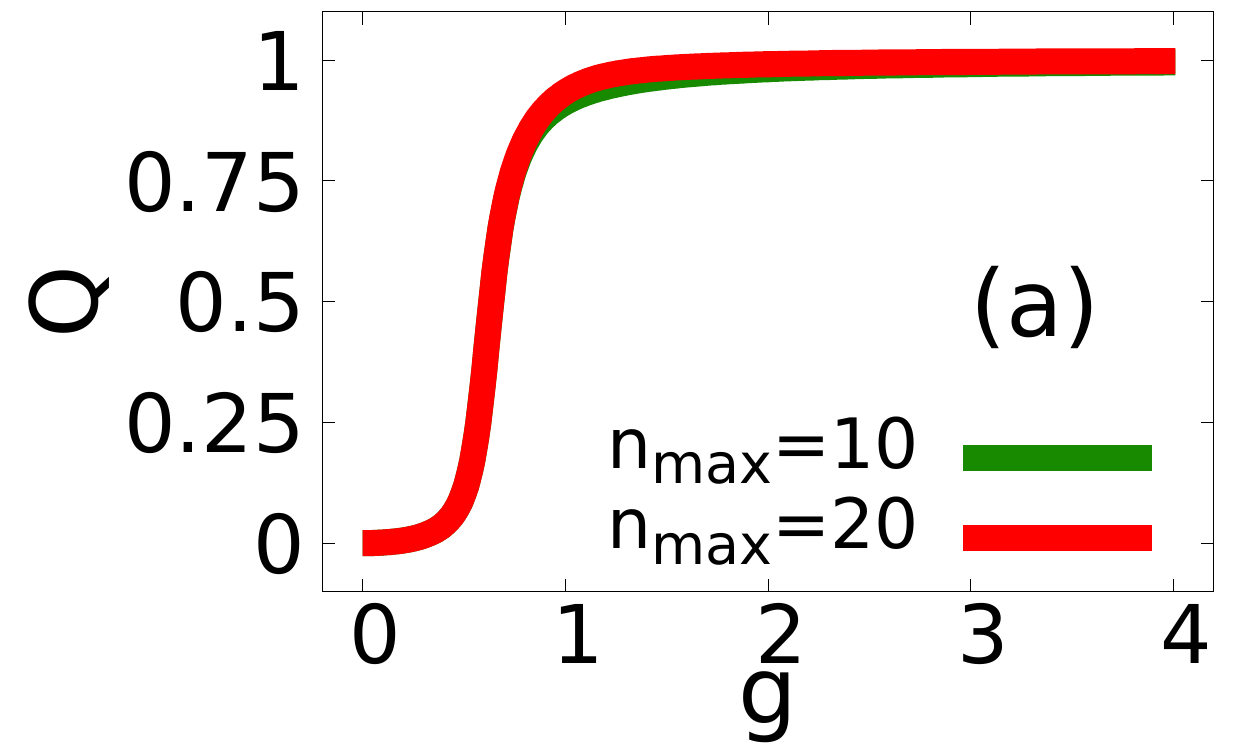}}\label{fig:sfig1}
      \subfigure{\includegraphics[width=0.32\textwidth]{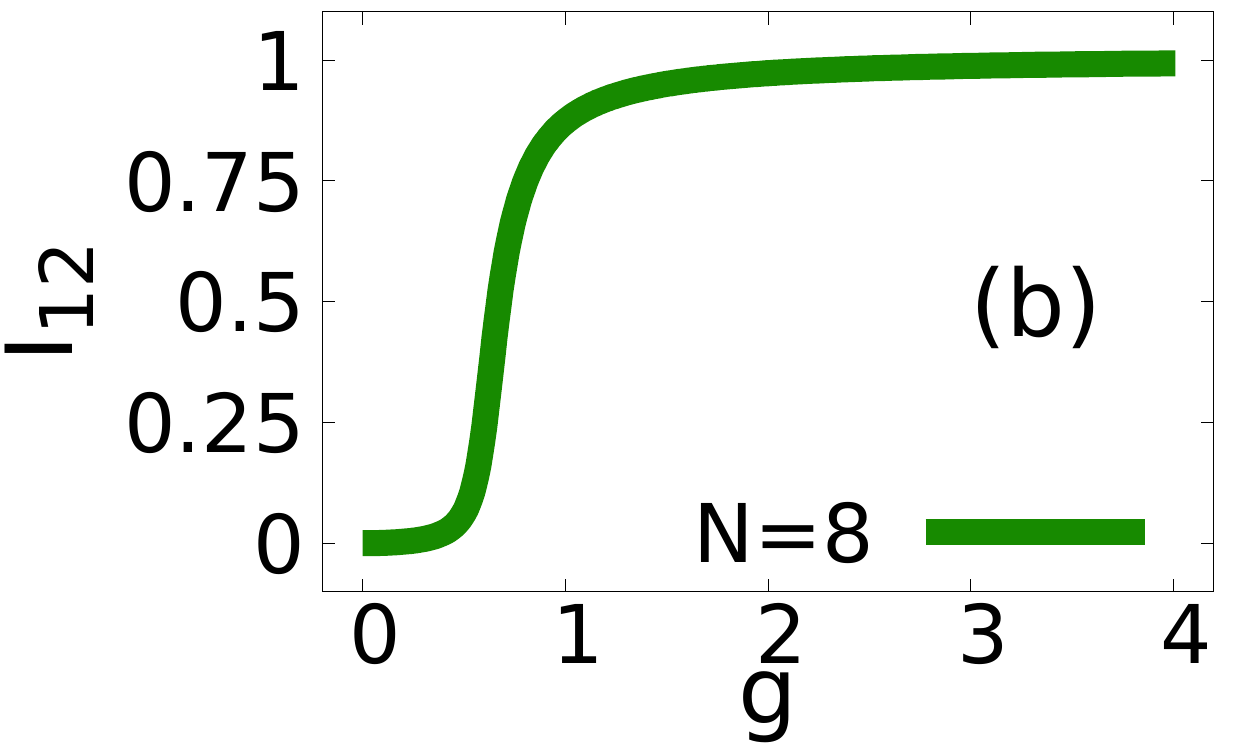}}\label{fig:sfig2}
      \subfigure{\includegraphics[width=0.32\textwidth]{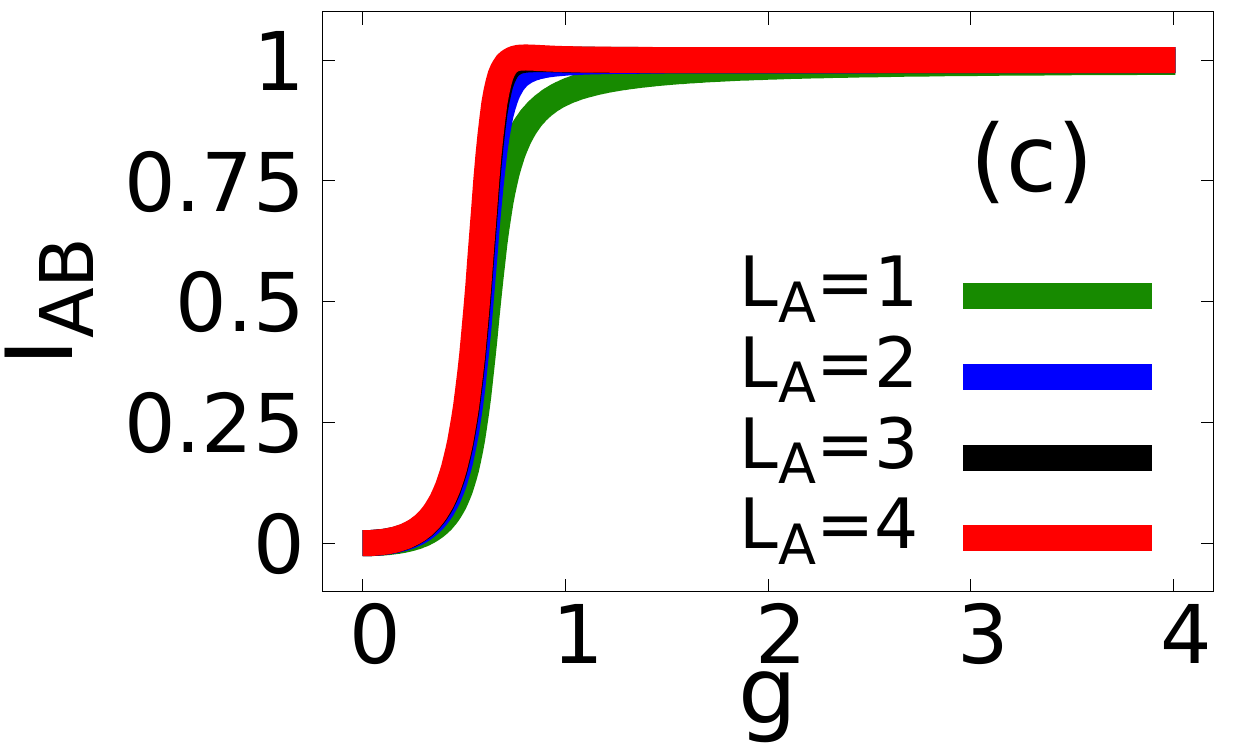}}\label{fig:sfig3}
      \caption{(a) The Meyer and Wallach Q measure of the ground state
        as a function of coupling, $g$. Number of spins $N=8$. (b)
        Mutual information between two spins. (c) Mutual information
        between two spin sub-sectors. Here we divide the spins into
        two sub-sectors with spin number $L_A$ and $L_B$. The total
        number of spins can be broken into $N=8=L_A + L_B = 1+7 = 2+6
        = 3+5 = 4+4$. For (b) and (c) $n_{max}=10$.}
      \label{fig:qpt1}
    \end{figure*}
    The so-called Meyer and Wallach $Q$
    measure~\cite{karthik2007entanglement, scott2004multipartite,
      radgohar2018global, bhosale2017signatures} defined as:
    \begin{equation}
    Q = 2\left[ 1 - Tr(\rho_1^{2}) \right],
    \end{equation}
    provides a useful signature at the QPT. In Fig.~\ref{fig:qpt1}(a)
    we exhibit the $Q$-measure for the ground state of the system as a
    function of coupling parameter $g$. Here $\rho_1$ is the single
    atomic reduced density matrix and it can be calculated by tracing
    out the bosonic part first and then over the $N-1$ atoms. $Q$ is
    close to zero for $g<g_c$ (NP) and close to one for $g>g_c$ (SP)
    with a transition near the critical point. It goes to zero when
    $Tr(\rho_1^{2}) \approx 1$ i.e., in the NP the single atom reduced
    density matrix behaves like a pure state having the two
    eigenvalues close to $1$ and $0$. On the other hand $Q \approx 1$,
    when $Tr(\rho_1^{2}) \approx 0.5 < 1$, hence $\rho_1$ is a mixed
    state in the SP.

    The mutual information $I_{12}$ between two spins $S_1$ and $S_2$ is given by
    \begin{align}
      I_{12} &= S_1 + S_2 - S_{12}\\
      S_{1,2} &= -Tr(\rho_{1,2} \ln(\rho_{1,2}))\nonumber\\
      S_{12} &= -Tr(\rho_{12}\ln(\rho_{12}))\nonumber
    \end{align}
    where $\rho_{1}$, $\rho_{2}$ are the reduced density matrices for
    the single spins and $S_{1}$, $S_{2}$ the corresponding von
    Neumann entropies, while $\rho_{12}$ is the reduced density matrix
    of two spins and $S_{12}$ is the corresponding entropy.
    Fig.~\ref{fig:qpt1}(b) shows the mutual information between two
    spins of the Dicke model. $I_{12}$ is close to zero in the NP and
    close to one in the SP with a transition near the critical
    point. Hence we can say that the total correlation between two
    spins is zero in the NP whereas in the SP the correlation is
    maximum. Thus $I_{12}$ significantly depends on the spin boson
    coupling $g$.  Fig.~\ref{fig:qpt1}(c) shows the mutual information
    between two groups of spins A and B, in which group A contains
    $L_A$ spins while group B contains the remaining $L_B=N-L_A$
    spins. Because of the symmetry in the coupling it does not matter
    which $L_A$ spins are considered. We see that this quantity too
    shows similar behaviour as $I_{12}$.  From Fig.~\ref{fig:qpt1}(a)
    we conclude that in the NP each spin is separately in a pure
    state, so there are no quantum correlations at all. On the other
    hand in the SP, we see that all the correlations are very high.
        
\subsection{Dynamics}
    \begin{figure*}[htbp]
      \subfigure{\includegraphics[width=0.33\textwidth]{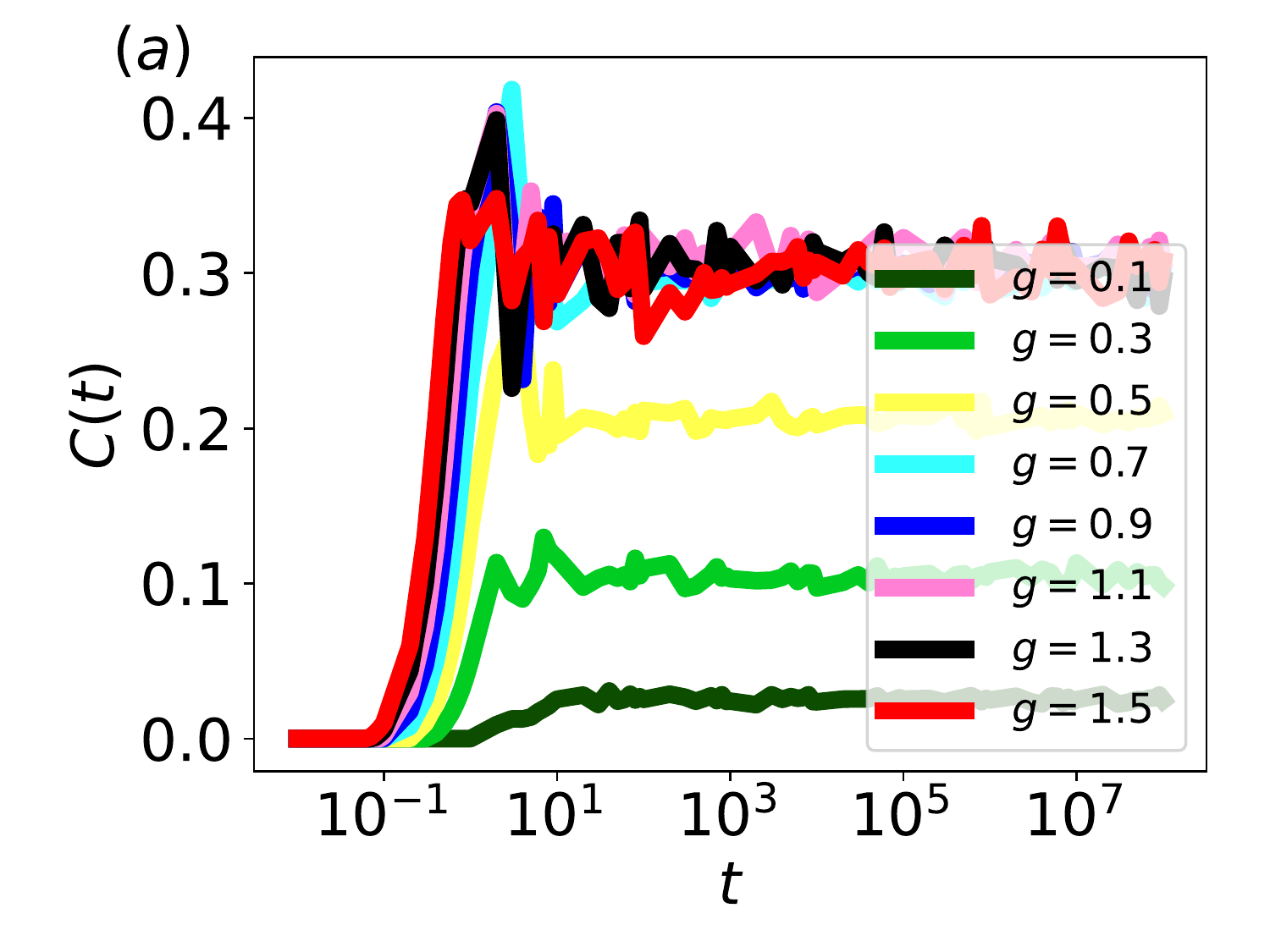}}\label{fig:sfig1}
      \subfigure{\includegraphics[width=0.33\textwidth]{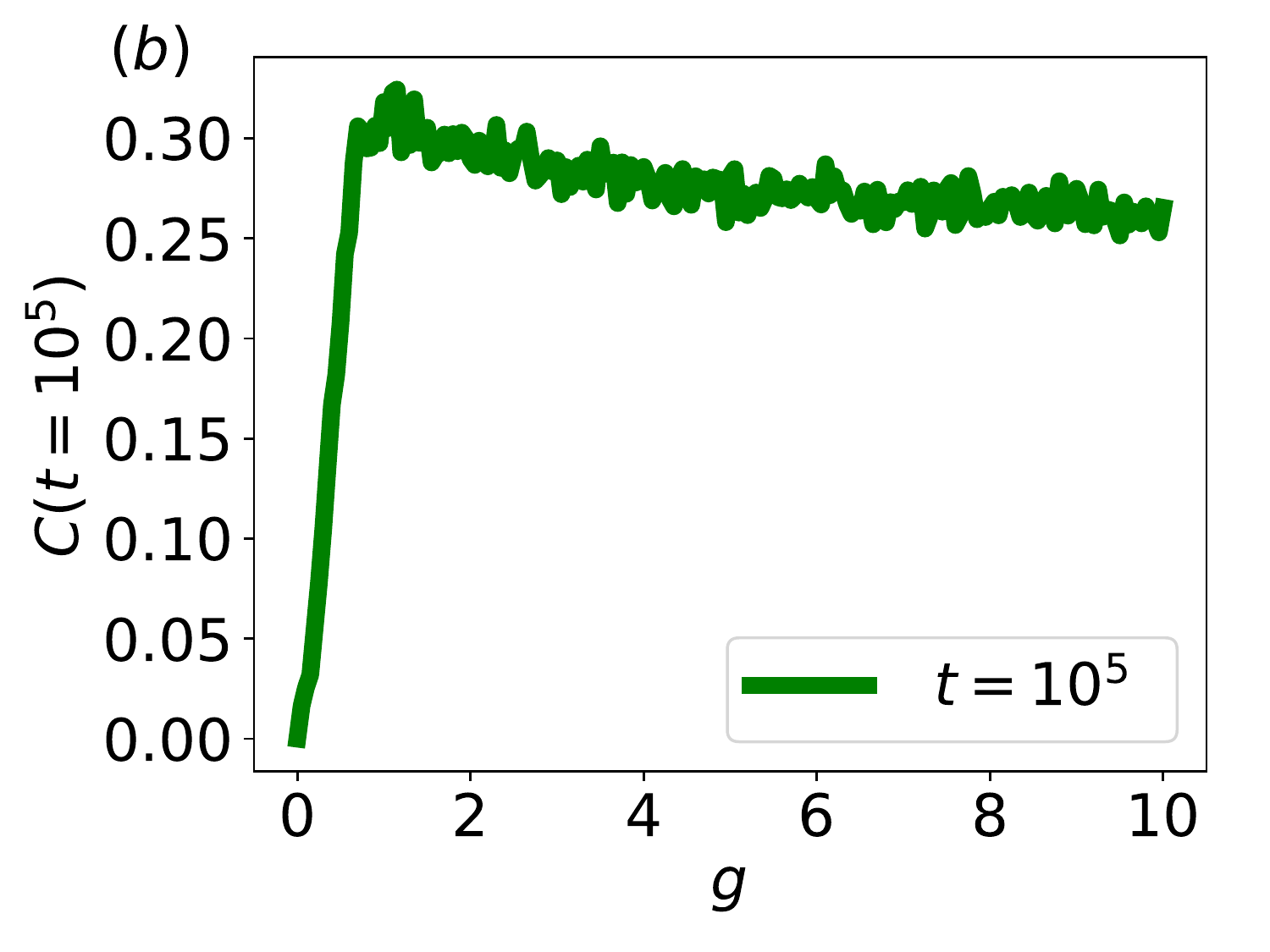}}\label{fig:sfig2}
      \subfigure{\includegraphics[width=0.305\textwidth]{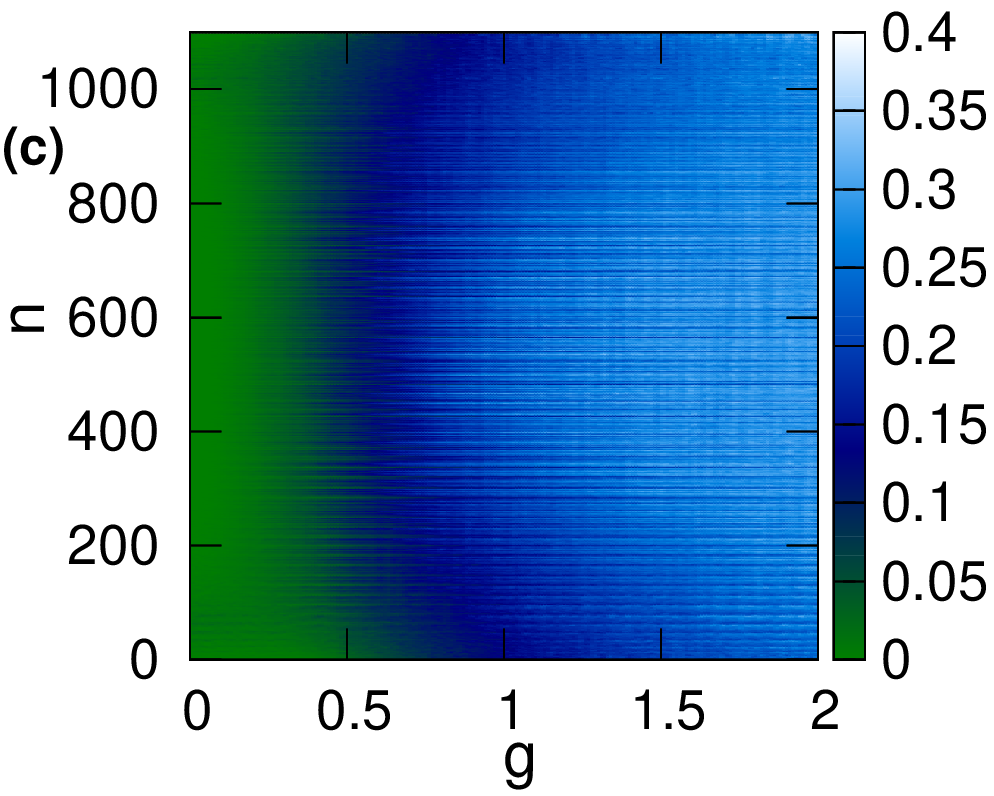}}\label{fig:sfig3}    
      \caption{(a) Concurrence between two atoms as a function of time
        $C(t)$ for different coupling strengths. (b) Saturation value
        of concurrence at time $t=10^5$ as a function of coupling
        strength $g$. The parameters for (a) and (b) are $N=128$,
        $n_{max}=32$. (c) Quench dynamics of concurrence $C$, at time
        $t=10^6$. $C(t=10^6)$ as a function of coupling parameter $g$
        and $n$ (index for the eigenstates of the decoupled
        Hamiltonian $\mathcal{H}_0$ accordingly). Green color denotes
        NP and blue color denotes SP. Parameters are $N=128$,
        $n_{max}=16$. }
      \label{fig:qd1}
    \end{figure*}

    We next describe how the quantum phase transition in the Dicke
    model is usefully studied with a quenching protocol. In a quantum
    quench, we prepare a closed system in an eigenstate of one
    Hamiltonian $\mathcal{H}_0$ and then have the system evolve
    dynamically in time under a different Hamiltonian $\mathcal{H} =
    \mathcal{H}_0 + \mathcal{H}_1$. For our problem, we take
    $\mathcal{H}_0$ to be the Dicke model with $g = 0$, and the middle
    excited state of $\mathcal{H}_0$ as our initial state. The value
    of $g$ is then suddenly changed to some other non-zero value of
    $g$, and the resulting dynamics of the system is studied.  The
    concurrence (between any two atoms) as a function of time $C(t)$
    is shown for a range of values of $g$ in Fig.~\ref{fig:qd1}(a).
    In general, $C(t)$ starting from zero remains close to zero upto
    $t\approx 0.1$, after which it increases upto $t\approx 10$. With
    further increase of time, it tends to saturate to a constant with
    some fluctuations. This saturating value of $C(t)$ depends on $g$,
    and carries information of the quantum phase transition. For
    $g=0.1$ (dark green color) the saturation value of concurrence is
    close to zero, but with increasing $g$, the saturation value
    increases gradually until a certain critical $g_c$. Beyond $g_c$
    (yollow color) we begin to see a saturating tendency of the
    saturation value itself with some fluctuations [see
      Fig.~\ref{fig:qd1}(a)]. This feature is made clearer by a plot
    of the saturation value of the concurrence at some late time, say
    $t=10^5$ as a function of the coupling parameter $g$ as shown in
    Fig.~\ref{fig:qd1}(b). We observe that in the NP the saturation
    value is increasing with $g$, but after crossing the critical
    point $g_c$, in the SP it is almost independent of $g$ i.e. the
    value is constant with some fluctuations. Thus we see that the
    dynamics too shows signatures of the quantum phase transition. In
    Fig.~\ref{fig:qd1}(c) we study the dynamics starting from each of
    the eigenstates of the decoupled Hamiltonian, $\mathcal{H}_0 =
    \omega a^{\dagger}a + \omega_0 J_z$, and evolving with the unitary
    time evolution operator $U=e^{-i\mathcal{H}t}$ where $\mathcal{H}$
    is the system Hamiltonian. For a particular coupling strength $g$,
    $U$ is fixed but the initial states are changing hence ${\vert
      \psi_{i}(t)\rangle} = U\vert {\psi_{i}(0)\rangle} = U\vert
    i\rangle$, where $\{ \vert i\rangle \}$ are the eigenstates of
    $\mathcal{H}_0$. We plot the concurrence (between any pair of
    atoms) wrt the time evolved. We see that Fig.~\ref{fig:qd1}(c)
    shows that the dynamics is sensitive to the quantum phase
    transition no matter which eigenstate we start from.

    \section{Thermal phase transition}
    We now show how the analytical expression for the transition
    temperature may be obtained. Beginning with the scaled Hamiltonian
    described by Eqn.(4) in the main Letter, we write the partition
    function of the Dicke model as:
    \begin{align}
      Z(N,T) = \sum_{s_1,..., s_N=\pm 1}\int \frac{d^2\alpha}{\pi}\langle s_1...s_N\vert\langle\alpha\vert e^{-\beta\tilde{\mathcal{H}}}\vert\alpha\rangle\vert s_1...s_N\rangle.
    \end{align}
    The expectation value of the Hamiltonian with respect to the
    bosonic modes is obtained in a straightforward manner:
    \begin{align}
      \langle\alpha\vert \tilde{\mathcal{H}}\vert\alpha\rangle = \alpha^*\alpha + \sum_{j=1}^N\Big[ \frac{\epsilon}{2}\sigma_j^z + \frac{\lambda}{\sqrt{N}}( \alpha + \alpha^{*} )\sigma_j^x\Big].
    \end{align}
    Defining
    \begin{align}
      h_j = \frac{\epsilon}{2}\sigma_j^z + \frac{\lambda}{\sqrt{N}}( \alpha + \alpha^{*} )\sigma_j^x
    \end{align}
    the expectation value with respect to the spins becomes a product
    of single-spin expectation values:
    \begin{align}
      \langle s_1...s_N\vert\langle\alpha\vert e^{-\beta \tilde{\mathcal{H}}}\vert\alpha\rangle\vert s_1...s_N\rangle = e^{-\beta\vert\alpha\vert^2}\Pi_{j=1}^N\langle s_j\vert e^{-\beta h_j}\vert s_j\rangle.
    \end{align}
    Using the above results, we can recast the partition function 
    in the form of a double integral as:
    \begin{align}
      Z(N,T) &= \int\frac{d^2\alpha}{\pi}e^{-\beta\vert\alpha\vert^2}\Big[\text{Tr}e^{-\beta h} \Big]^N\\
      &= \int\frac{d^2\alpha}{\pi}e^{-\beta\vert\alpha\vert^2}\Big(2\cosh\Big[ \frac{\beta\epsilon}{2}\Big[ 1 + \frac{16{\lambda}^2\alpha^2}{\epsilon^2 N} \Big]^{1/2} \Big]\Big)^N.\nonumber
    \end{align}    
    Here $\alpha$ is real, so $\vert\alpha\vert = \vert\alpha^{*}\vert$. Now
    \begin{align}
      \int\frac{d^2\alpha}{\pi} = \int_0^{\infty}r d r\int_0^{2\pi}\frac{d\theta}{\pi} = 2\int_0^{\infty} r dr.
    \end{align}
    Defining $y=\frac{r^2}{N}$ and 
    \begin{align}
      \phi(y) = -\beta y + \ln\Big( 2\cosh\Big[ \frac{\beta\epsilon}{2}\Big[ 1 + \frac{16{\lambda}^2 y}{\epsilon^2} \Big]^{1/2} \Big] \Big),
    \end{align}
    we have:
    \begin{align}
      Z(N,T) = N \int_0^{\infty} d y\exp\Big( N \phi(y)  \Big).
    \end{align}
    Since we are interested in the thermodynamic limit where
    $N\to\infty$, we can invoke Laplace's
    method~\cite{jeffreys1999methods}, and the integral is given by
    \begin{align}
      Z(N,T) = N\frac{C}{\sqrt{N}}\max_{0\leq y\leq \infty}\exp\Big( N\Big[ \phi(y) \Big] \Big)
    \end{align} 
    where $C$ is some constant. To find the maximum of the function $\phi(y)$, we compute its derivative:
    \begin{align}
      \phi^{\prime} = -\beta + \frac{\beta 4{\lambda}^2}{\epsilon}\frac{1}{\eta}\tanh\Big( \frac{\beta\epsilon\eta}{2} \Big)
    \end{align} 
    where
    \begin{align}
      \eta = \Big[ 1 + \frac{16{\lambda}^2 y}{\epsilon^2} \Big]^{1/2}.
      \label{eqn:eta}
    \end{align}  
    Putting
    \begin{equation}
      \phi^{\prime} = 0,
    \end{equation}
    we get
    \begin{align}
      \eta = \frac{4{\lambda}^2}{\epsilon}\tanh\Big( \frac{\beta\epsilon\eta}{2} \Big).
      \label{eq8}  
    \end{align} 
    The hyperbolic tangent funtion is a monotonically increasing
    function and is bounded above by unity. Since $\eta \ge 1$ by
    definition (Eqn.~\ref{eqn:eta}), if $4{\lambda}^2<\epsilon$, there is no
    solution for Eqn.~\ref{eq8}. On the other hand, for
    $4{\lambda}^2>\epsilon$, the solution depends on the value of $\beta$.
    The critical value of the inverse temperature $\beta_c$ can be
    computed by putting $\eta = 1$ and is given by:
    \begin{align}
      \beta_c = \frac{2}{\epsilon}\tanh^{-1}\Big( \frac{\epsilon}{4{\lambda}^2} \Big).
    \end{align}
    Thus substituting $\epsilon=\frac{\omega_0}{\omega}$ and $\lambda=\frac{g}{\omega}$, 
    we have an exact expression for the transition temperature:
    \begin{equation}
      T_c = \frac{1}{\beta_c} = \Big(\frac{\omega_0}{2\omega}\Big)\frac{1}{\tanh^{-1}\Big( \frac{\omega\omega_0}{4 g^2} \Big)}.
      \label{eqn:Tc}
    \end{equation}

\section{Excited state quantum phase transition (ESQPT)}
\begin{figure}[htbp]
      \subfigure{\includegraphics[width=0.33\textwidth]{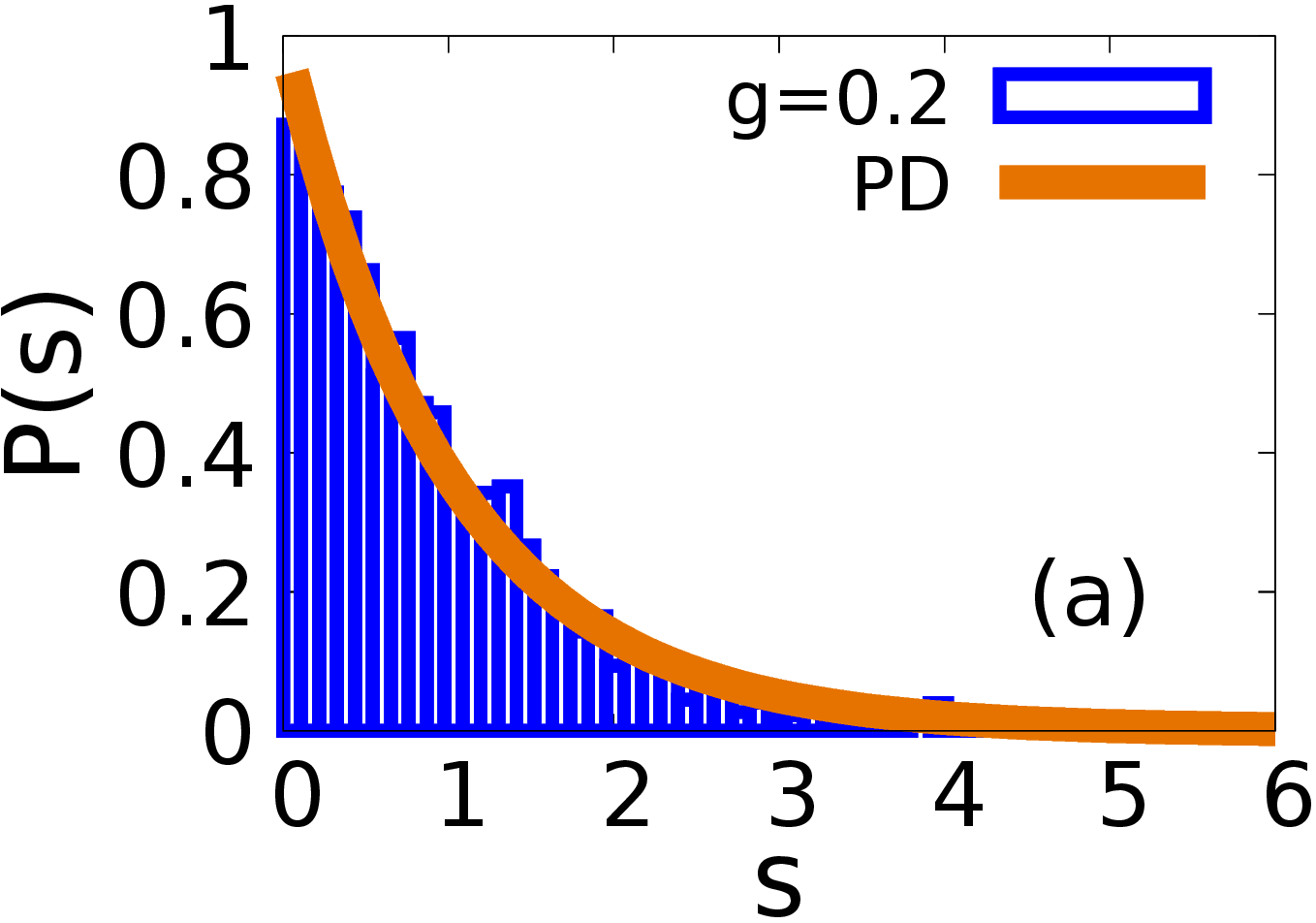}}\label{fig:sfig1}
      \subfigure{\includegraphics[width=0.33\textwidth]{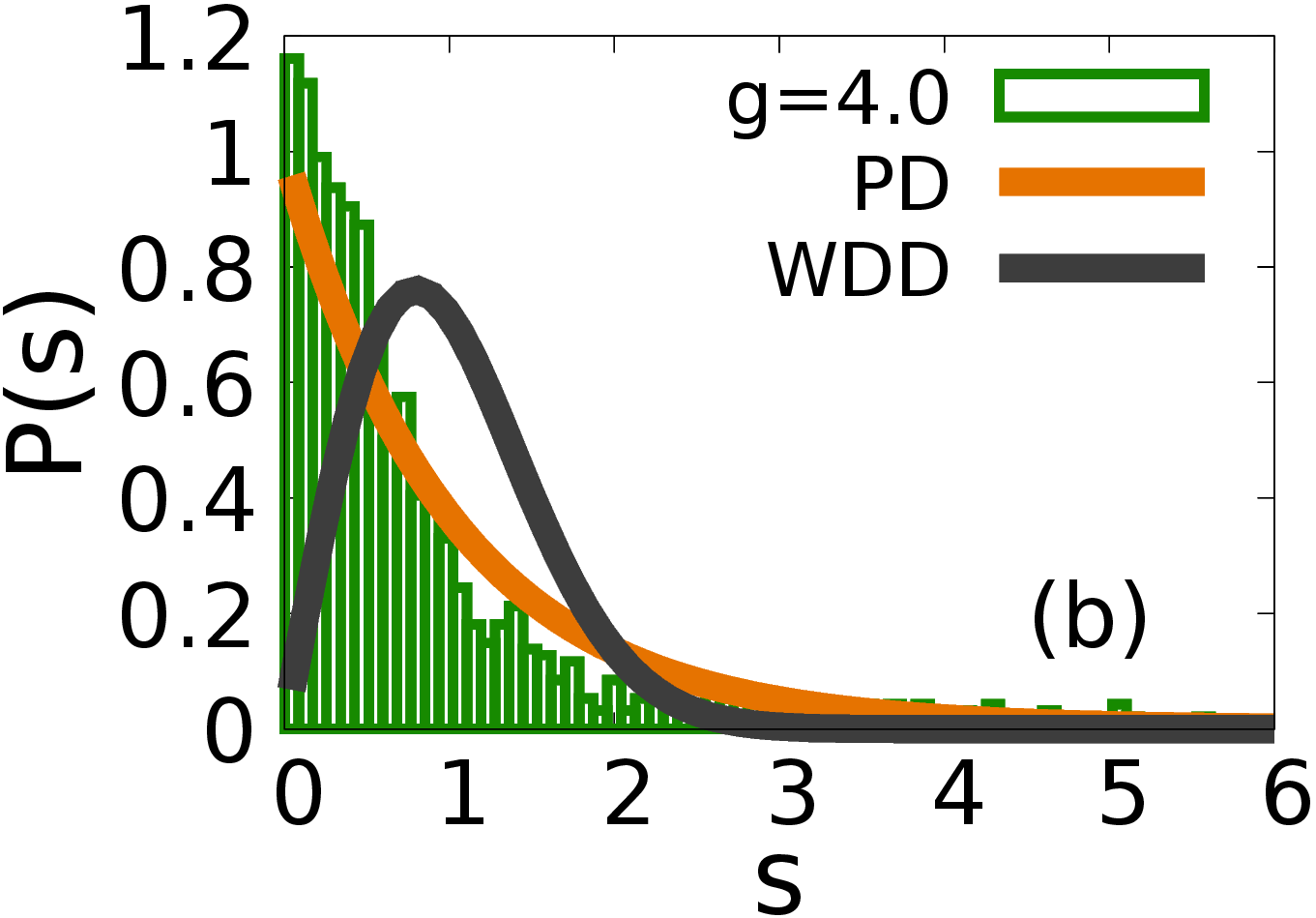}}\label{fig:sfig2}
      \subfigure{\includegraphics[width=0.33\textwidth]{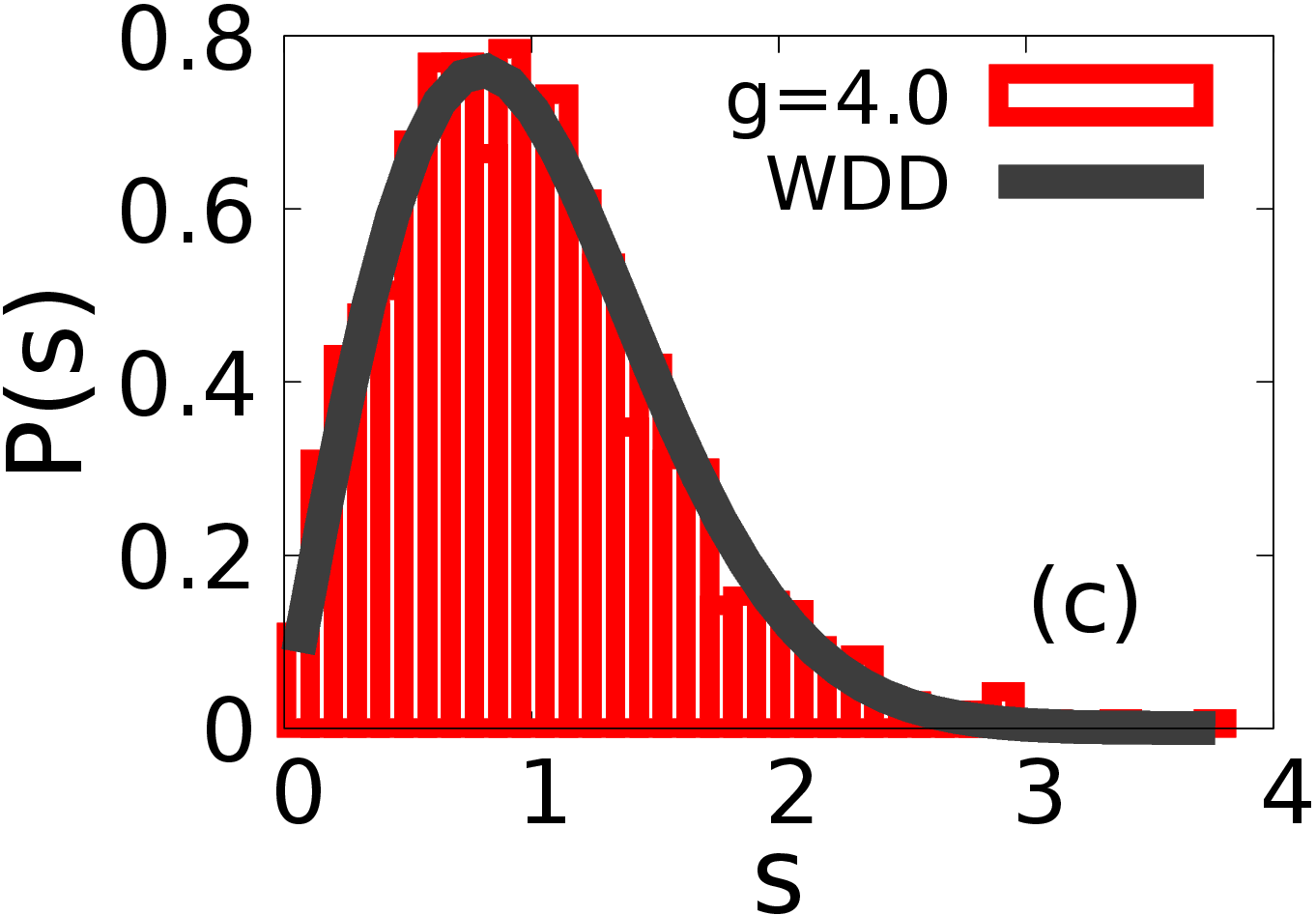}}\label{fig:sfig3}
      \subfigure{\includegraphics[width=0.33\textwidth]{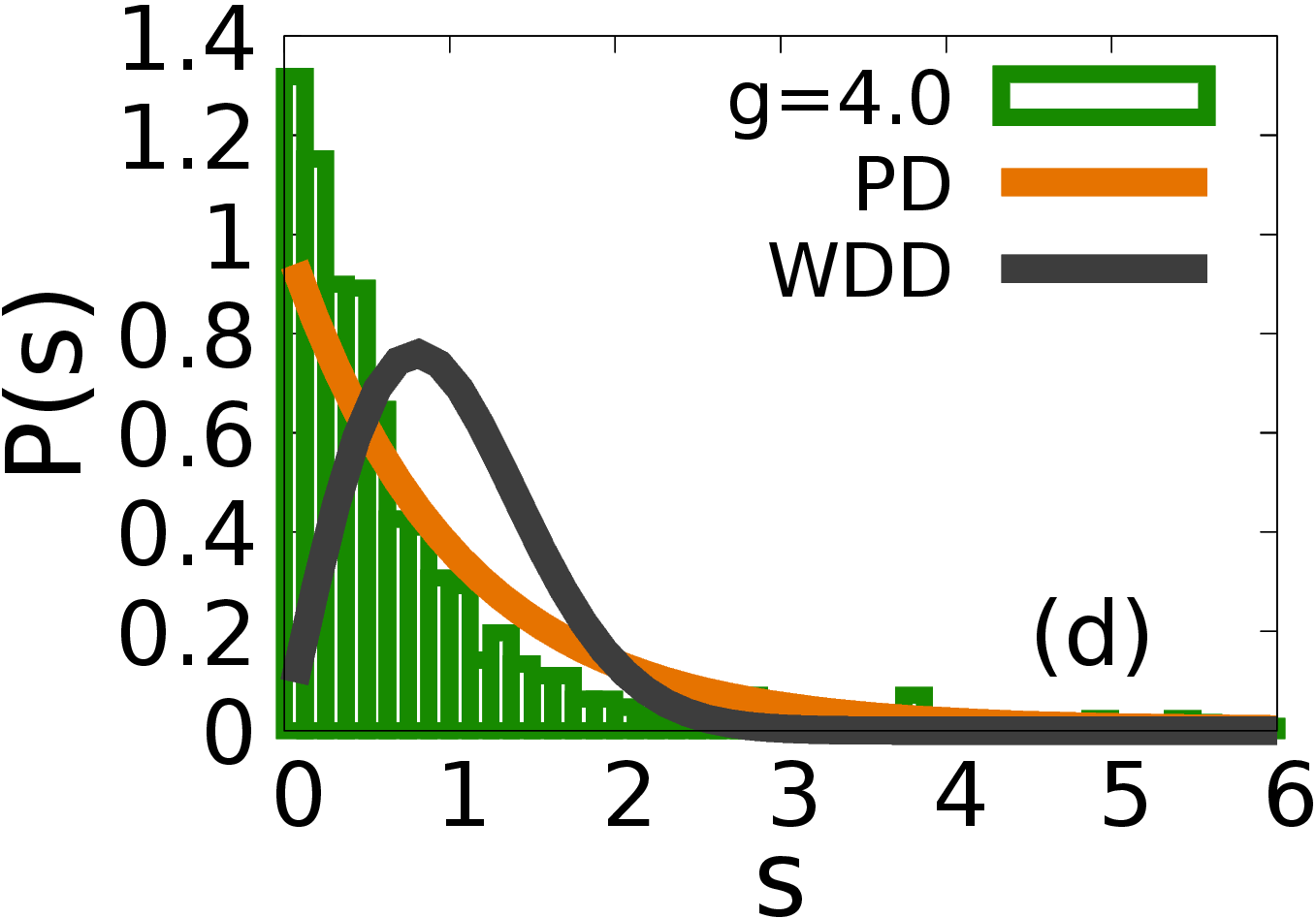}}\label{fig:sfig4}    
      \caption{Level statistics of the spectrum. (a) For $g<g_c$ the
        level spacing distribution is more like Poisonian. For $g>g_c$
        there are three bands in the spectrum: below the lower cut-off
        energy (b), above the upper cut-off energy (d), and the
        energies in between (c). The energy spacing distribution in
        between the lower and upper cut-off is more like
        Wigner-Dyson. However in (b), (d) the behaviours are not
        clear, as they show mixed behaviour, although they are closer
        to Poissonian behaviour.}
      \label{fig:esqpt3}
    \end{figure}
        
    The Dicke model exhibits an excited state quantum phase transition
    in the super-radiant phase. When $g> g_c$, while it is well
    known~\cite{ perez2011excited, lewis2019unifying} that the
    eigenvalues above a cut-off energy $E_c$ behave in a distinctly
    different manner in comparison with the eigenvalues below the
    cut-off, we report that in fact there is not just a lower cut-off,
    but also an upper cut-off.  While the main Letter describes
    several properties connected to eigenstates, here we show
    \emph{eigenvalue} data that suggest that the central band of
    energy levels between a lower and upper cut-off behave differently
    from the lower and upper energy bands.
    \begin{figure}[htbp]
      \includegraphics[width=0.4\textwidth]{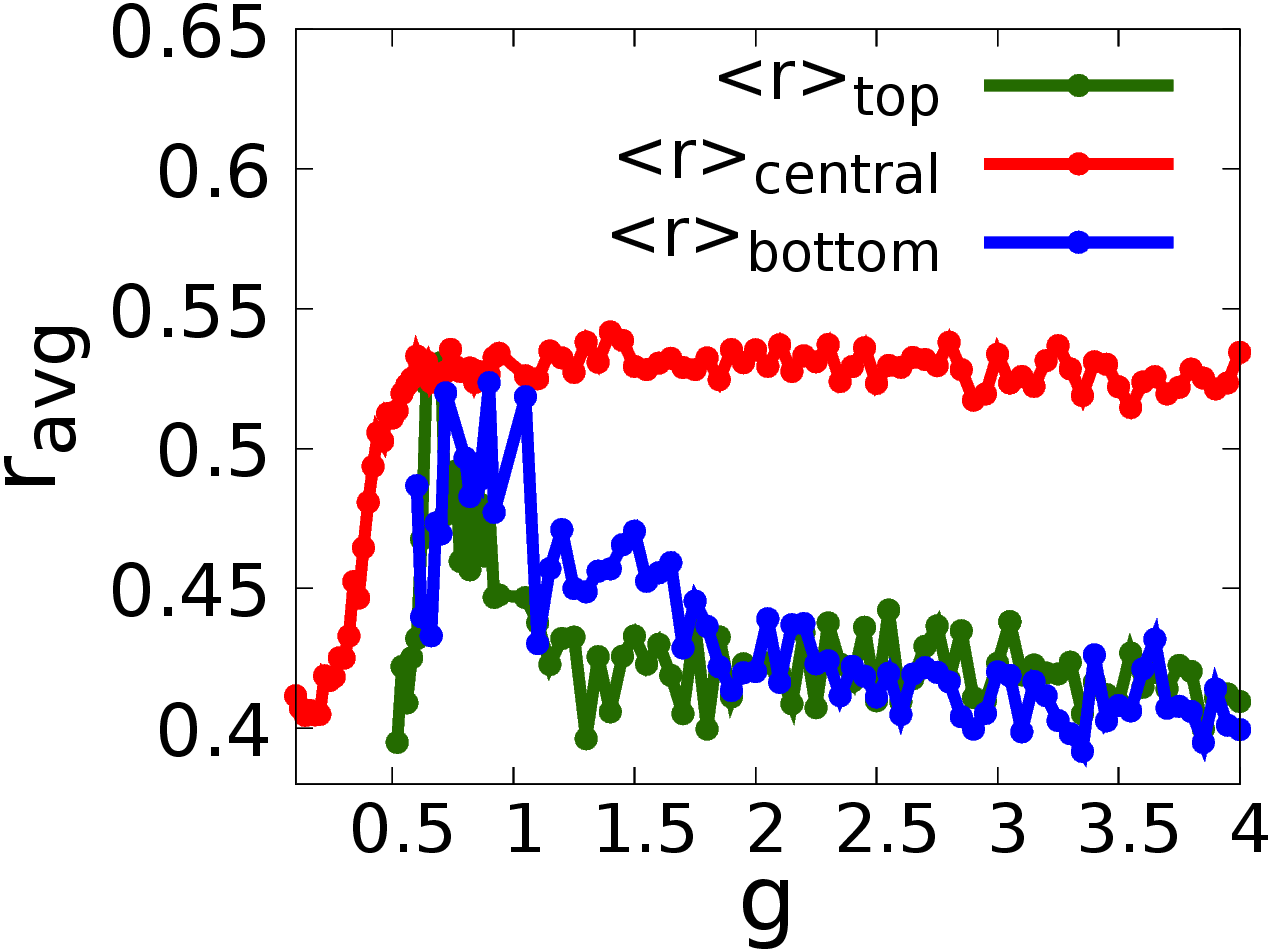}
      \caption{$r_{avg}$ value as a function of coupling, for the different energy bands. $(i)$ Blue color is used for energies below the lower cutoff, $(ii)$ green color for energies above the upper cutoff, and $(iii)$ red color for the middle band energies. Parameters are $n_{max}=200$, $N=60$.}
      \label{fig:esqpt4}
    \end{figure}

    The onset of ergodic behaviour is typically diagnosed by inspection
    of the level spacing distribution~\cite{poilblanc1993poisson}. Let
    $\{ E_n \}$ denote the energy levels of the DM in ascending
    order. Under the assumption that the density of states equals
    unity, the distribution $P(s)$ of the level spacings $s_n =
    E_{n+1} - E_n$ is given by the Poisson distribution $P(s) =
    \exp(-s)$ in the normal phase~\cite{berry1977level}. On the other
    hand, in the super-radiant phase, the level spacings adhere to the
    Wigner-Dyson distribution $P(s) = \frac{\pi}{2}s\exp\Big[
      -(\pi/4)s^2 \Big]$~\cite{bohigas1984characterization}. To study
    the level statistics we consider two $g$ values: $g=0.2<g_c$ and
    $g=4.0>g_c$. For $g<g_c$ we see that the energy spacings are
    consistent with the Poisson distribution (PD). For $g>g_c$ we
    study the level statistics separately in three bands as shown in
    Fig.~\ref{fig:esqpt3}. While the energy spacing distribution for
    levels that lie between the lower and upper cut-off energies is
    like the Wigner-Dyson distribution, the level spacing
    distributions of the upper and lower energy bands show mixed
    behaviour, although the distribution looks more Poissonian than
    Wigner-Dyson. Evidently, there is a striking absence of level
    repulsion in these bands, in stark contrast to the levels in the
    central band. While the presence of the lower cut-off is reported
    in the literature~\cite{ perez2011excited, lewis2019unifying}, our
    data clearly reveal an upper cut-off as well.
    
    The above picture with respect to the energy levels is further
    strengthened by a study of the ratio of consecutive level
    spacings, which has now become a standard
    measure~\cite{atas2013distribution}. Let ${s_n}$ denote level
    spacing between two consecutive energies $E_{n+1}$ and $E_n$. The
    average spacing ratio $\langle r \rangle$ is defined as the
    average over $n$ of the ratio of consecutive level spacings:
    \begin{equation}
      r_{n} =
      \frac{\text{min}(s_{n-1},\hspace{1mm}s_{n})}{\text{max}(s_{n-1},\hspace{1mm} s_n)}.
    \end{equation}
    From random matrix theory it is known
    that~\cite{atas2013distribution} $\langle r \rangle$ takes a value
    $\langle r \rangle\approx0.386$ for quasi-integrable Hamiltonians
    and $\langle r \rangle\approx0.5307 $ for Hamiltonians from the
    Gaussian orthogonal ensemble (GOE). For $g<g_c$ $\langle r
    \rangle\approx0.386$ and for $g>g_c$ $\langle r
    \rangle\approx0.5307 $ for the central band. For the upper and the
    lower energy bands $\langle r \rangle$ lies in between $0.386$ and
    $0.5307$. However, we observe that as the coupling strength
    increases, and as the atomic number becomes large, they tend to
    resemble the Poisonian ensemble [see Fig.~\ref{fig:esqpt4}].
